\documentclass[preprint,12pt]{elsarticle}

\usepackage{amssymb}
\usepackage{epsfig}
\usepackage{url}
\usepackage{colordvi}
\usepackage{txfonts}
\usepackage{graphicx}
\usepackage{bm}

\usepackage{xspace,url}

\journal{J. Comp. Phys.}

\def \pd{{\partial}}
\def\nut{\nu}
\def\Rey{{\rm Re}}
\def\Le{{\rm Le}}

\newcommand{\SSSS}{\mbox{\boldmath ${\sf S}$} {}}
\newcommand{\FF}{\mbox{\boldmath $F$} {}}
\newcommand{\UU}{\mbox{\boldmath $U$} {}}
\newcommand{\JJ}{\mbox{\boldmath $J$} {}}
\newcommand{\VV}{\mbox{\boldmath $V$} {}}
\newcommand{\qq}{\mbox{\boldmath $q$} {}}
\newcommand{\Eq}[1]{Eq.~(\ref{#1})}
\newcommand{\Eqs}[2]{Eqs.~(\ref{#1}) and~(\ref{#2})}

\def\nab{{\bm{\nabla}}}
\def\ff{{\bm f}}

\def\VV{{\bm V}}

\newcommand{\EQ}{\begin{equation}}
\newcommand{\EN}{\end{equation}}
\newcommand{\EQA}{\begin{eqnarray}}
\newcommand{\ENA}{\end{eqnarray}}
\newcommand{\bez}{\begin{eqnarray*}}
\newcommand{\eez}{\end{eqnarray*}}
\newcommand{\be}{\begin{equation}}
\newcommand{\ee}{\end{equation}}
\newcommand{\beq}{\begin{eqnarray}}
\newcommand{\eeq}{\end{eqnarray}}
\newcommand{\bc}{\begin{center}}
\newcommand{\ec}{\end{center}}

\begin{document}

\begin{frontmatter}



\title{A high-order public domain code for direct numerical simulations
of turbulent combustion}


\author{N.\ Babkovskaia$^1$, N.\ E.\ L.\ Haugen$^2$, A.\ Brandenburg$^{3,4}$}

\address{
$^1$Division of Geophysics and Astronomy
(P.O. Box 64), FI-00014 University of Helsinki, Finland\\
$^2$SINTEF Energy Research, Kolbj{\o}rn Hejes vei 1a, N-7465 Trondheim, Norway\\
$^3$NORDITA, AlbaNova University Center, Roslagstullsbacken 23,
SE-10691 Stockholm, Sweden\\
$^4$Department of Astronomy, Stockholm University, SE-10691 Stockholm, Sweden
}

\begin{abstract}
  A high-order scheme for direct numerical simulations
  of turbulent combustion is discussed.
  Its implementation in the massively parallel and publicly available
  {\sc Pencil Code} is validated with the focus on hydrogen combustion.
  Ignition delay times (0D) and laminar flame velocities (1D) are
  calculated and compared with results from the commercially available
  Chemkin code.
  The scheme is verified to be fifth order in space.
  Upon doubling the resolution, a 32-fold increase in the accuracy of
  the flame front is demonstrated.
  Finally, also turbulent and spherical flame front velocities are calculated
  and the implementation of the non-reflecting so-called
  Navier-Stokes Characteristic Boundary Condition is validated
  in all three directions.
\end{abstract}


\end{frontmatter}



\section{Introduction}
%\label{}

Modeling of turbulence is one of the largest research areas within flow 
mechanics.
Turbulent combustion inherits all the properties of non-reacting turbulent flow.
The most important addition is linked to the highly nonlinear reaction processes,
and models for this are called combustion models. Two additional challenges in 
turbulent combustion are the very sharp changes in density and differential 
diffusion of mass and heat.

For combustion processes it is crucial to be able to simulate the mixing of the 
combustible species correctly. Traditionally this has been done by means
of mixing models in Reynolds Averaged Navier Stokes (RANS) codes by 
combining, e.g., the $k$-$\epsilon$ turbulence model and 
the eddy dissipation concept (or EDC) ``mixing'' model \citep{MH76,M81,M89},
or in Large Eddy Simulation (LES) \citep{PV05} 
where a sub-grid model is used both for the turbulence and 
for the scalar mixing. There are however major and still unresolved
problems related to modelling of what happens on the very 
smallest scales with these methods.

Several RANS codes with detailed chemistry are commercially available
\citep{c10,f10}, and there are a huge number of these codes found as in-house
codes at different academic institutions and in many industrial departments 
around the world.
There are also freely available open-source RANS codes with 
detailed chemistry \citep{o10}.
The reason for the popularity of RANS is its low demand on computational
resources. Because of this RANS has, for decades, been 
the most used type of code for industrial purposes.

Nevertheless, also LES has increased in popularity during the last years,
and this has led, for example, to the 
inclusion of a LES module in \cite{f10}. Most LES codes for combustion
today are, however, in-house codes owned by different academic institutions.

The most accurate way of simulating turbulent combustion is
to use Direct Numerical Simulation (DNS)~\citep{PV05} 
instead of RANS or LES.   
In DNS one resolves the full range of time and length scales 
of both the turbulence and the chemistry (using accurate high-order
numerical methods for computational efficiency). 
The problem with DNS 
is however that it is very resource demanding, both on CPU-hours and memory. 

In this paper we present the implementation of a detailed chemistry
module in a finite-difference code \citep{PC} for compressible
hydrodynamic flows.
The code advances the equations in non-conservative form.
The degree of conservation of mass, momentum and energy can then be used to
assess the accuracy of the solution.
The code uses six-order centered finite differences. 
For turbulence calculation we normally use the RK3-2N scheme of \cite{Wil80}
for the time advancement \citep{BD02}.
This scheme is of Runge-Kutta type, third order, and it uses only two chunks
of memory for each dependent variable.
For hydrodynamic calculations,
the lengths of the time step is calculated based on a number of constraints
involving maximum values of velocity, viscosity, and other quantities
on the right-hand sides of the evolution equations.
In some cases we use instead a fifth-oder Runge-Kutta-Fehlberg scheme
with an automatic adaptive time step, subject to the aforementioned
hydrodynamic constraints.
However, in many cases we found it advantageous to use a fixed
time step whose length is estimated based on earlier trial runs
with an automatically calculated time step.

On a typical processor, the cache memory between the CPU and the RAM
is not big enough to hold full three-dimensional data arrays.
Therefore, the {\sc Pencil Code} has been designed to evaluate first all
the terms on the right-hand sides of the evolution equations along a
one-dimensional subset (pencil) before going to the next pencil.
This implies that all derived quantities exist only along pencils.
Only in exceptional cases do we allocate full three-dimensional arrays
to keep derived quantities in memory.
However, most of the time, multiple operations including the calculation
of derivatives is performed without using intermediate storage.

As far as we are aware, no open source high-order DNS code
with detailed chemistry is currently available.
The amount of man-hours for implementing a fully parallelized
DNS code with detailed chemistry is enormous.
It is therefore now timely to make such a code available in the public
domain and to encourage further development by a wider range of scientists.
Here we describe the implementation of such a scheme in the {\sc Pencil Code},
which is currently maintained under the Google Code subversion repository,
\url{http://pencil-code.googlecode.com/}.
The code is highly modular and comes with a large selection of physics modules.
It is portable to all commonly used architectures using Unix or Linux
operating systems.
The code is well documented and independent of external libraries and
any third party licenses.
All parts of the code, including the current chemistry implementation,
is therefore explicitly open source code.
In particular, there are no pre-compiled binary files. 
Consequently there are no licenses required for running any part of the code.
It is therefore straightforward to download the full source code from the
original subversion repository on google-code.
The Message Passing Interface libraries are needed when running on
multiple processors, but all parts of the code can also run on a
single processor without these libraries.
The integrity of the code is monitored through the automatic execution
of a selection of test cases on various platforms at different sites.
The detailed history of the code with about 14,000 revisions is accessible.

It should be emphasized that the use of high-order discretization
is critical for optimizing the accuracy at a given resolution.
Doubling the resolution of a 3D explicit code require 16 times more CPU time,
but this increases the accuracy by a factor of 32.
In fact, switching to a derivative module with a tenth order scheme is
straightforward and not significantly more expensive.

\section{The equations}
In this section we present the governing equations together with
the required constituent relations such as the equation of state and expressions
for viscosity, diffusivity and conductivity. 

\subsection{Governing equations}
The continuity equation is solved in the form
\begin{eqnarray}
\frac{{\rm D}\ln \rho}{{\rm D} t} = -\nab \cdot \bm{U},
\label{eq:rho}
\end{eqnarray}
where ${\rm D}/{\rm D}t= \pd/\pd t + {\bm U}\cdot {\bm \nabla}$ is the 
advective derivative,
$\rho$ is the density, and ${\bm U}$ is the velocity.
The momentum equation is written in the form
\begin{eqnarray}
\frac{{\rm D} {\UU}}{{\rm D} t}= \frac{1}{\rho} \left( -\nab p +{\bm F_{vs}}
\right) + \ff,
\label{eq:UU}
\end{eqnarray}
where $p$ is pressure,
$\ff$ is a volume force (e.g.\ gravity or a random forcing function),
\EQ
\label{F_vs}
\FF_{\rm vs}={\bm \nabla}\cdot(2\rho\nut \SSSS)
\EN
is the viscous force, where
${\sf S}_{ij}={1\over2}(\partial U_i/\partial x_j+\partial U_j/\partial x_i)
-{1\over3}\delta_{ij}\nabla\cdot\bm{U}$
is the trace-less rate of strain tensor.
The equation for the mass fractions of each species is
\begin{equation}
\rho \frac{{\rm D} Y_k}{{\rm D} t}  =
-\nabla \cdot {\JJ_k} + \dot{\omega}_k,
\label{eq:Y}
\end{equation}
where $Y$ is the mass fraction,
$\JJ$ is the diffusive flux, $\dot{\omega}$ is the reaction rate and 
subscript $k$ refers to species number $k$.
Finally, the energy equation is
\begin{eqnarray}
\left( c_p - \frac{R}{m}\right) \frac{{\rm D} \ln T}{{\rm D} t}  =
\sum_k \frac{{\rm D} Y_k}{{\rm D} t} \left(\frac{R}{m_k}- \frac{h_k}{T} \right) -\frac{R}{m} \nabla \cdot {\bm U} +\frac{2 \nut \SSSS^2}{T} -\frac{\nabla \cdot {\bf q}}{\rho T}, \label{eq:energy}
\end{eqnarray}
where $T$ is the temperature,
$c_p$ is the heat capacity at constant pressure,
$R$ is the universal gas constant,
$h$ is the enthalpy,
$m$ is the molar mass,
and $\qq$ is the heat flux.
The reason for solving for the temperature directly, instead of, e.g., the
total energy, is to avoid having to find the temperature from
the total energy afterwards.
In this work we use the ideal gas equation state given by
\EQ
p=\frac{\rho R T}{m}.
\EN
In the following we discuss the detailed expressions for viscosity,
reaction rate, species diffusion, thermal conduction,
enthalpy and heat capacity.

\subsection{Viscosity}

The viscosity $\nu$ is the viscosity of the mixture given by \citep{w50}
\be
\nu=\sum_{k=1}^{N_s} \frac{X_k \nu_k}{\sum_{j=1}^{N_s} X_j \Phi_{kj}},
\ee
where $N_s$ is the number of species, 
$\nu_k$ is the single component viscosity,
$X_k=Y_k m/m_k$ is the mole fraction of species $k$,  
and
\be
\Phi_{kj}=\frac{1}{\sqrt 8} \left(1 + \frac{m_k}{m_j} \right)^{-1/2}
\left\{
1+\left(\frac{\nu_k}{\nu_j}\right)^{1/2} \left(\frac{m_j}{m_k}\right)^{1/4} \right\}^2.
\ee
The single component viscosity is given as \citep{ch81}
\be
\nu_k=\frac{5}{16} \frac{\sqrt{\pi k_B T m_k}}{\pi \sigma^2_{k} \Omega^{(2,2)*}_k},
\ee
where $\sigma_k$ is the Lennard-Jones collision diameter,
$k_B$ is the Boltzmann constant, and $\Omega^{(2,2)*}_k$ is
the collision integral that is given by \citep{MR77}
\be
\Omega^{(2,2)*}_k=\Omega^{(2,2)*}_{{\rm L-J},k}+\frac{0.2\delta^*_k}{T^*_k},
\ee
where $\Omega^{(2,2)*}_{{\rm L-J},k}$ is the Lennard-Jones collision integral
and
\be
\delta^*_k=\frac{\mu_k^2}{2\epsilon_k \sigma_k^3}, 
\;\;\;
T^*_k=\frac{k_B T}{\epsilon_k}
\ee
are the reduced dipole moment and temperature, respectively.
In the above equations,
$\epsilon_k$ is the Lennard-Jones potential well depth and
$\mu_k$ is the dipole moment.

The values of $\epsilon_k$, $\mu_k$ and $\sigma_k$ must be given as input 
\citep{MM61},
while the Lennard-Jones collision integral is represented by
\be
\label{omega_22}
\Omega^{(2,2)*}_{{\rm L-J},k}=\left[\sum_{i=0}^7 a^{(2)}_i (\ln T^*_k)^i\right]^{-1},
\ee
where the coefficients $a^{(2)}_i$ are found from Table~\ref{ai}.

\begin{table}
\begin{center}
\begin{tabular}{|l|r|r|}
\hline
$i$ & \multicolumn{1}{c|}{$a^{(1)}_i$}
    & \multicolumn{1}{c}{$a^{(2)}_i$} \\
\hline
$0$ &$ 6.96945701\times 10^{-1}$ &$ 6.33225679\times 10^{-1}$ \\
$1$ &$ 3.39628861\times 10^{-1}$&$ 3.14473541\times 10^{-1}$ \\
$2$ &$ 1.32575555\times 10^{-2}$ &$ 1.78229325\times 10^{-2}$ \\
$3$ &$-3.41509659\times 10^{-2}$ &$-3.99489493\times 10^{-2}$ \\
$4$ &$ 7.71359429\times 10^{-3}$ &$ 8.98483088\times 10^{-3}$ \\
$5$ &$ 6.16106168\times 10^{-4}$ &$ 7.00167217\times 10^{-4}$ \\
$6$ &$-3.27101257\times 10^{-4}$ &$-3.82733808\times 10^{-4}$ \\
$7$ &$ 2.51567029\times 10^{-5}$ &$ 2.97208112\times 10^{-5}$ \\
\hline
\end{tabular}
\caption{
The $a_i$ coefficients are used in \Eqs{omega_22}{omega_11}
and are taken from the paper of \cite{e07}.
\label{ai}}
\end{center}
\end{table}

\subsection{Reaction rate}
The reaction rate of species $k$ is given by
\beq
\dot \omega_k=m_k
\sum_{s=1}^{N_r}(\nu^{\prime\prime}_{ks}-\nu^{\prime}_{ks}) \left[
\left(\frac{\rho_k}{m_k}\right)^{\sum_{i=1}^{N_s}(\nu^{\prime}_{ki})}k_s^+\prod_{j=1}^{N_s}X_j^{\nu_{js}^{\prime}}
-\left(\frac{\rho_k}{m_k}\right)^{\sum_{i=1}^{N_s}(\nu^{\prime\prime}_{ki})}k_s^-\prod_{j=1}^{N_s}X_j^{\nu_{js}^{\prime\prime}}\right],
 \eeq
where $N_r$ is the number of chemical reactions, 
$m_k$ is the molar mass of species $k$,
$p_k$ is the partial pressure of species $k$, 
$n_k=\rho_k/m_k$ is the molar concentration of species $k$, and 
$\rho_k$ is the density of species $k$. 
Furthermore,
$\nu^{\prime}_{ks}$ and $\nu^{\prime\prime}_{ks}$ are the stoichiometric 
coefficients of species $k$ of reaction $s$
on the reactant and product side, respectively.
The rates of reaction $s$ are given by the Arrhenius expression
\EQ
k_s=B_n T^{\alpha_n}\exp(-E_{an}/RT), 
\EN
where $B_n$ is the pre-exponential factor, 
$\alpha_n$ is the temperature exponent, 
and $E_{an}$ is the activation energy and they are all empirical coefficients that
are given by the kinetic mechanism. For hydrogen-air combustion,
an example of a kinetic mechanism is found in \cite{li04}.

\subsection{Species diffusion}
The diffusion flux is $\JJ_k=\rho Y_k \VV_k$.
Following \cite{w85}, the diffusion velocity, $\VV_k$,  is found by solving
\EQ
\label{multicomp}
\nab X_p 
= \sum_{k=1}^{N_s} \frac{X_p X_k}{D_{pk}}\left(\VV_k-\VV_p\right)
+ \left(Y_p-X_p\right) \frac{\nab p}{p}
+ \frac{\rho}{p}\sum_{k=1}^{N_s} Y_pY_k(\ff_p-\ff_k),
\EN 
where the Soret effect is neglected. The first term on the right hand side
corresponds to ordinary diffusion, the second term is the so called
baro-diffusion, while the last term is due to unequal body-forces per unit mass
among the species.
Unfortunately the CPU cost of solving \Eq{multicomp} numerically scales as
$N_s^2$ for each grid point and time-step, and simplifications are therefore
required in order to be able to run reasonably sized simulations.
 
In the mixture averaged approximation the diffusion velocity is expressed 
as \citep{h69}
\EQ
 \label{V_k}
 \VV_k=-\frac{D_k {\bm d}_k }{X_k}, \quad
 {\bm d}_k=\nab X_k + (X_k -Y_k) \frac{1}{p} \nab p,
\EN
where the body force term has been neglected,
$D_k$ is the diffusion coefficient for species $k$
\be
D_k=\frac{1-Y_k}{\sum^{N_s}_{j \ne k} X_j/D_{jk}},
\label{eq:Dk}
\ee
and $D_{kj}$ is the binary diffusion coefficient that is given by \citep{h69}
\be
D_{kj}=\frac{3}{16} \frac{\sqrt{2 \pi k_B^3 T^3/m_{jk}}}
{P \pi \sigma^2_{jk} \Omega^{(1,1)*}_{jk}},
\ee
where $\sigma_{jk}=(\sigma_j + \sigma_k)/2$ is the reduced collision diameter,
$m_{jk}$ is the reduced molecular mass for the $(j,k)$ species pair
\be
m_{jk}=\frac{m_j m_k}{m_j+m_k},
\ee
$\Omega^{(1,1)*}$ is the collision integral that is given by \citep{e07}
\be
\label{omega_11}
\Omega^{(1,1)*}_{jk}=\Omega^{(1,1)*}_{\rm L-J}+\frac{0.19 \delta_{jk}^*}{T^*_{jk}},
\quad\quad
\Omega^{(1,1)*}_{\rm L-J}=\left[\sum_{i=0}^7 a^{(1)}_i (\ln T^*)^i\right]^{-1},
\ee
where the coefficients $a^{(1)}_i$ are also found from Table~\ref{ai}.
The reduced dipole moment and the reduced temperature are given by 
\EQ
\delta_{jk}^*=\frac{1}{2} \mu_{jk}^{*2}\quad\mbox{and}\quad
T_{jk}^*=\frac{k_B T}{\epsilon_{jk}},
\EN
respectively, where $\mu_{jk}^{*2}= \mu_{j}^*\mu_{k}^*$
is the nondimensional 2-species dipole moment,
$\epsilon_{jk}=\sqrt{\epsilon_j \epsilon_k}$
is the 2-species Lennard-Jones potential, and
$\mu_k^*=\mu_k/\sqrt{\epsilon_k \sigma_k^3}$
is the nondimensional dipole moment.

\subsection{Thermal conduction}
The heat flux is given by
\EQ
\qq= \sum_k h_k \JJ_k - \lambda \nab T,
\EN
where $\lambda$ is the thermal conductivity, which
is found from the thermal conductivities of the
individual species as
\be
\label{eq:lambda}
\lambda= \frac{1}{2} \left(
\sum_{k=1}^{N_s} X_k \lambda_k + \frac{1}{\sum_{k=1}^{N_s}X_k/\lambda_k}
\right).
\ee
Here, the individual species conductivities are composed of transitional, 
rotational and vibrational contributions and are given by \citep{Wa92}
\be
\lambda_k=\frac{\nu_k}{m_k}(f_{trans.}C_{v,trans.}+f_{rot.}C_{v,rot.}
+f_{vib.}C_{v,vib.} ).
\ee

\subsection{Enthalpy and heat capacity}
The enthalpy of the ideal gas mixture can be expressed in terms of 
isobaric specific heat $c_p$ and temperature as 
\begin{eqnarray}
h_i=h_i^0+\int_{T_0}^T c_{p,i} {\rm d}T, \;\;\;
h=\sum_{i=1}^{N_s} Y_i h_i,
\label{eq:hh}
\end{eqnarray}
where $h_i^0$ is the enthalpy of formation of species $i$ at temperature $T_0$.

To calculate the heat capacity $c_p$ we use a Taylor expansion,
\be
c_p=\frac{R}{m} \sum_{i=1}^{5} a_i T^{i-1},
\label{eq:cp}
\ee
where $a_i$ are coefficients found in \cite{GM71}.

\section{Scaling in the Pencil Code}

\subsection{General remarks}

For direct numerical simulations (DNS) it is crucial to have high accuracy.
This is due to the fact that we are interested in resolving the
smallest scales, and consequently we can not allow for these scales to 
be lost due to low accuracy. Furthermore, for many situations it is
important to know the actual Reynolds number of the simulation. 
The Reynolds number is defined as
\EQ
%\Rey=\frac{ul}{\nu_{\rm eff}},
%AB: this discussion about effective seems unclear to me.
%AB: I think we can just skip it.
\Rey=\frac{ul}{\nu},
\EN
where $u$ and $l$ are characteristic velocity and length scale, 
respectively, and $\nu$ is the viscosity.
High accuracy is obtained by the use of high order discretization.
In the {\sc Pencil Code}, sixth order discretization is normally used
\citep{B03}.
However, for the density a fifth order upwinding scheme us used.

\subsection{One-step reaction model, $R \rightarrow P$.}
\label{sec:1step}
In order to verify that the code recovers correct scaling,
we use simplified chemistry and compare against known results.
Following \citet{DH07}, we consider a one-step laminar premixed 
flame model. The irreversible reaction can be presented as $R \rightarrow P$, 
where $R$ is the reactant and $P$ is a product. Using the approach of 
\citet{FE93}, we neglect viscous effects, and take $\rho$, 
$\lambda$, $C_p$ and the $x$-component of the velocity to be constant.
Then the  system of equations takes the form
\begin{eqnarray}
\frac{\partial Y_p}{\partial t} + U_x \frac{\partial Y_p}{\partial x} &=&
 \frac{1}{Le} \frac{\partial^2 \phi}{\partial x^2}   + \dot \Omega,
\\
\frac{\partial \ln T}{\partial t} + U_x \frac{\partial \ln T}{\partial x}&=& 
 -\frac{\rho U_x^2 C_p\beta (\beta-1) }{\lambda}  \frac{(T_{\infty}-T_0)}{T} 
+ \frac{\lambda}{\rho C_p T} \frac{\partial^2 T}{\partial x^2},
\end{eqnarray} 
where the reaction rate is defined as 
\EQ
\dot \Omega =  \left\{
\begin{array}{ccl}
\rho U_x^2 C_p\lambda^{-1} \beta (\beta-1) (1- Y_p)& & \mbox{if $T>T_c$}\\
0                                                  & & \mbox{otherwise},  
\end{array}
\right.
\EN
where $\beta=(T_{\infty}-T_0)/(T_{\infty}-T_c)$, while $T_0$  and
 $T_{\infty}$ are the temperature of the unburned and burned gas, respectively, 
$T_c$  is the critical temperature, $Y_p$ is a mass fraction of the product, 
$\Le=\lambda/(\rho D C_p)$ is the Lewis number, $D$ is the mass diffusion 
coefficient. 
\begin{figure*}
\centerline{\epsfig{file=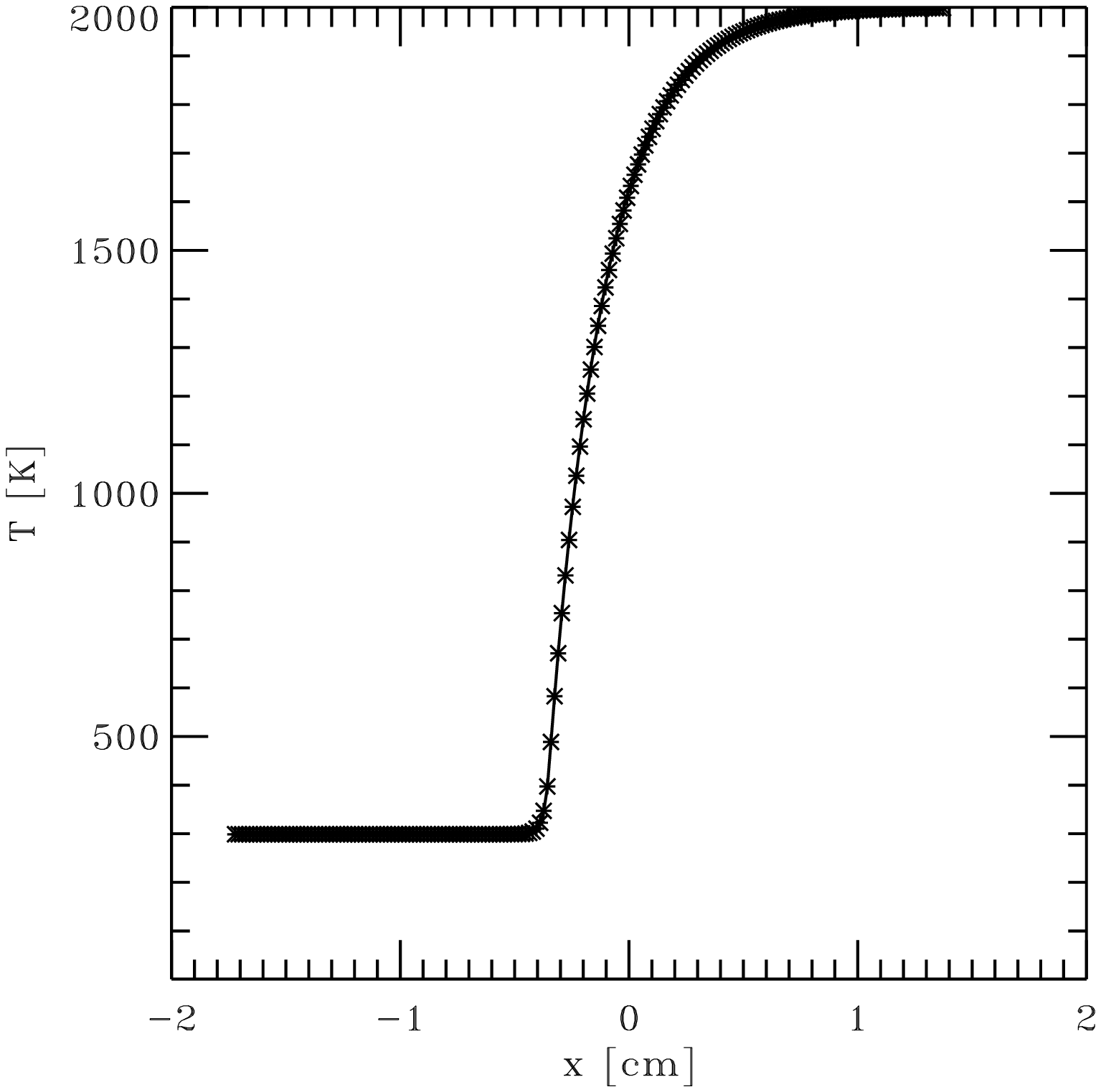,width=7cm}\epsfig{file=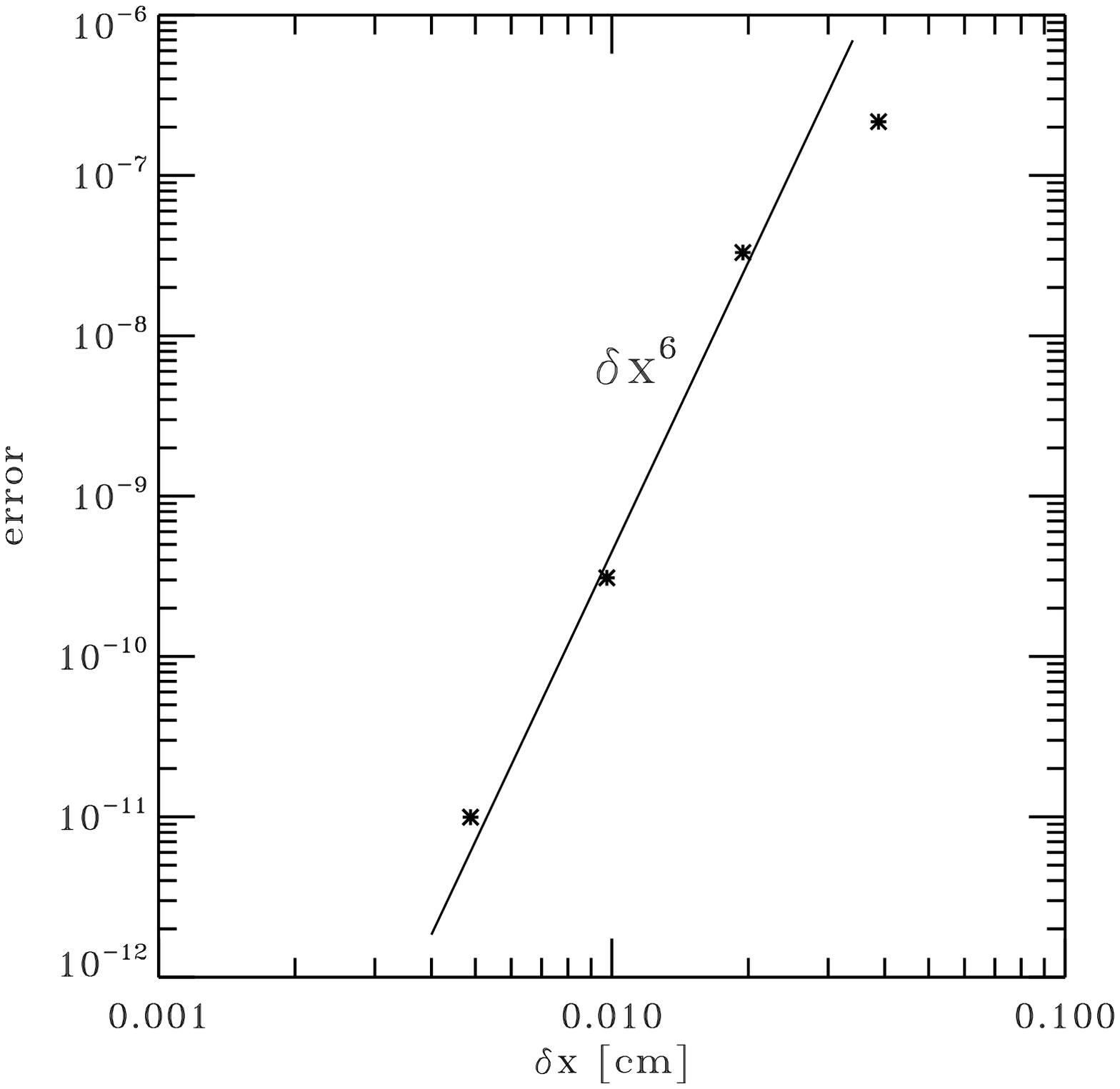,width=7cm}} 
\caption{One-step laminar premixed flame model.  {\rm \it Left panel}: 
temperature as a function of $x$ obtained numerically  
(solid curve)  and analytically (asterisks). 
{\rm \it Right panel}: error of the calculation
as a function of the mesh spacing $\delta x$ is shown by asterisks,
and the expected dependence of error (proportional to $\delta x^6$)
is indicated by the solid line.} 

\label{fig:T_an}
\end{figure*}
Taking $\Le=1$,  one obtains the following analytical solution
\begin{equation}
\tilde{T}=\left\{
\begin{array}{lcl}
1-\beta^{-1} \exp(x/\delta)           & & \mbox{   if  $x<0$,} \\
1-\beta^{-1} \exp[(1-\beta) x/\delta] & & \mbox{ otherwise,}
\end{array}
\right. 
\end{equation}
where $\tilde T={(T-T_0)}/(T_{\infty}-T_0)$ and $\delta=\lambda/(\rho U C_p)$ 
is a characteristic thickness.

In the left hand plot of Fig.~\ref{fig:T_an} we compare the numerical 
results with the analytical solution for 
$T_0=300$ K, $T_{\infty}=2000$ K, $T_c=440$ K, $\beta=1.09$,  
$C_p=10^8$ erg g$^{-1}$ K$^{-1}$, $\lambda=10^4$ erg cm$^{-1}$ 
K$^{-1}$ s$^{-1}$, $D=2$ cm$^2$ s$^{-1}$, $\rho=5 \times 10^{-4}$ 
g cm$^{-3}$ and $U_x=100$ cm $s^{-1}$. It can be seen that there is good 
agreement between the numerical and analytical results.
To show the high-order spatial accuracy provided by  the Pencil 
Code, we obtain the set of 
solutions for 33, 65, 129, 257, 513 and 1025 grid points, and compare 
them pairwise ("33" with "65", "65" with "129" and so on). In every pair
we compare only the points which are collocated, that is, we do the comparison
for all the grid points of the coarser grid against half of the grid points
of the finer grid.  
The time step is controlled by the chemistry and is here fixed at
$\delta t=10^{-8}\,{\rm s}$.
The size of the domain is 3 cm. The maximum absolute value of the difference 
between the corresponding solutions in common points is taken as the error. 
In the right-hand panel of Fig.~\ref{fig:T_an} the error as a function 
of $\delta x$ is shown by the symbols.  
One can see that sixth-order accuracy is obtained (see solid line).

\subsection{One-dimensional premixed flame with the Li mechanism}
\label{sec:1D}
\begin{figure}
\centerline{\epsfig{file=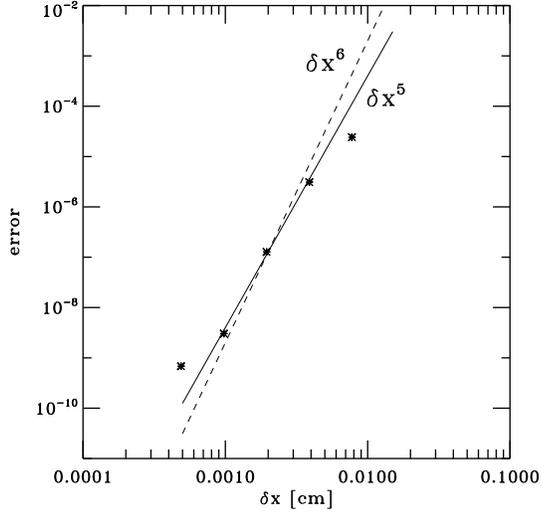,width=7cm}} 
\caption{
Accuracy of calculation as a function of $\delta x$ for the Li mechanism.
An error of calculations
as a function of the mesh spacing $\delta x$ is shown by asterisks,
and the solid line is the dependence of the error, which is
proportional to $\delta x^5$.}
\label{fig:T_Li}
\end{figure}
In this section we study a one-dimensional problem with detailed chemistry. 
We consider hydrogen-air combustion using the Li mechanism \citep{li04}.  
The fresh hydrogen-air mixture enters the domain under stoichiometric conditions
($Y_{\rm  H_2}=2.4$ $\%$, $Y_{\rm  O_2}=23$ $\%$ and $Y_{\rm N_2}=74.6$ $\%$)
at a temperature of $T_u=298 K$ and a pressure of $p=1$~atm. 
To avoid reflection of acoustic waves at the boundaries
non-reflecting boundary conditions
are required. Here the Navier-Stokes Characteristic Boundary Conditions (NSCBC) 
\citep{PL92,LD08} have been used. 

To check the spatial accuracy in the case of the Li mechanism,
we perform numerical experiment as described in  Sec.~\ref{sec:1step}
for  65, 129, 257, 513, 1025 and 2049 grid points.
The time step is fixed at $\delta t=10^{-10}$s. The error as a 
function of $\delta x$ is presented in Fig.~\ref{fig:T_Li},
where one can see that the fifth-order spatial accuracy is achieved. 
The reason we get only fifth order, and not sixth order, is that we are
using upwinding for the density, which has the effect of decreasing the 
order of the discretization to fifth order \citep{BD02}.

\section{Validation of the chemistry implementation in the Pencil Code}
In this section the chemistry module will be verified quantitatively by 
comparison with the commercially available simulation tool \cite{ch10}. 
In order to minimize the effect
of the fluid flow, and to focus as much as possible on the chemistry, these
tests have been restricted to zero and one dimensional tests cases.

\subsection{Zero-dimensional test: ignition delay}
\label{sec:0d}
\begin{figure}
\centerline{\epsfig{file=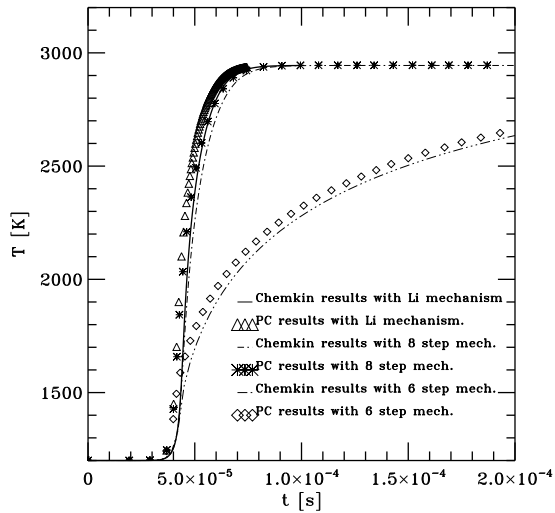,width=7cm}} \caption{Dependence of the gas 
temperature on time, computed with the 6-step mechanism ({\it diamonds}), 
the 8-step mechanism ({\it asterisks}), and the Li mechanism ({\it triangles}). 
The numerical results are compared with Chemkin for the 6-step mechanism 
({\it three doted-dashed curve}), for the 8-step mechanism ({\it doted-dashed 
curve}), and for the Li mechanism ({\it solid curve}).
 }
\label{fig:0d}
\end{figure}
First a zero-dimensional ignition delay test for different
chemical mechanisms is studied and compared with the results obtained 
with Chemkin for the same setup. 
One assumes hydrogen-air combustion in a closed homogeneous reactor at constant
volume, and consider the 6-step and 8-step mechanisms of 
\cite{SM07} together with the Li mechanism \citep{li04}. 
The initial values are $p=1$~atm for the pressure,
$\phi=1$ for the equivalence ratio, and $T=1200$~K for the temperature.
As the minimum time step varies greatly with the progress of the combustion
process the time step is here chosen automatically using the adaptive 
Runge-Kutta-Fehlberg method. 
The results are presented in Fig.~\ref{fig:0d}, where one can
see good agreement with the Chemkin results.

\subsection{One-dimensional test: laminar flame speed}
\label{sec:1D_flame}
Next, we consider a one-dimensional flame front.
The cold premixed gas enters at one end of the domain at given
velocity. Inside the domain there is a flame front where the fuel is 
consumed and the temperature increases to the mixture flame temperature.
The mechanism of \cite{li04} is used and the inlet values of
temperature, pressure 
and mixture compositions are the same as described in Sec.~\ref{sec:1D}. 
The inlet velocity is adjusted such that the flame front becomes stationary inside
the domain. The flame velocity is thus arranged to be equal to the inlet
velocity.

\begin{figure}
\centerline{\epsfig{file=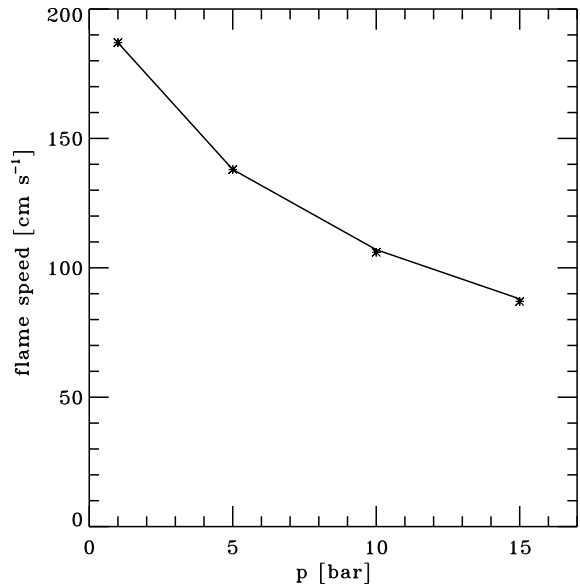,width=8cm}}
\caption{Flame speed velocity as a function of pressure $p$.
Chemkin results are shown by asterisks.
}\label{fig:Vf_p}
\end{figure}

We find that the flame front should be resolved by at least
10 grid points in order to ensure a well resolved flame. 
For a thickness of the flame front of about 0.01~cm, and
a domain of $\Delta x = 0.1$~cm, the optimal grid size is found to be 150
points.  
The flame speed as a function of pressure is shown in 
Fig.~\ref{fig:Vf_p} where the current results are found to compare well
with those of Chemkin. 

\section{Three-dimensional flame front simulations}

\subsection{Plane flame front}

In this section we study a 3D representation of the initially flat flame front. 
The settings of the problem is
similar to that in Section~\ref{sec:1D_flame}, i.e.\ initially 
the temperature, density and 
velocity change in the $x$ direction, and are constant
in the $y$ and $z$ directions.
We use periodic boundary conditions in the $y$ and $z$ directions,
and in the $x$ direction we use inlet and outlet NSCBC boundary conditions
on the left and right hand sides, respectively, as was done in \cite{LD08}.
The pressure is $p=1$ bar,  the initial gas temperature is $T=750$ K, and 
the inlet velocity is 30 m~s$^{-1}$.
The unburned gas mixture has an equivalence 
ratio of $\phi=0.8$.
The size of the calculated domain is taken to be
$0.5\times 0.25\times 0.25\,{\rm cm}^3$,
and the grid size is (128 $\times$ 64 $\times$ 64).

\begin{table}
\begin{center}
\begin{tabular}{lllll}
\hline
Run & Mechanism & Number of species & Transport & $\tau$ [$\mu s/N_t/N_g$] \\
\hline
A &   Li &   13&  Mix-aver.  &    71.3\\
B &   Li &   13&  Oran       &    51.7\\
C &   No &   13&  Mix-aver.  &    42.1\\
D &   No &   13&  Oran       &    24.3\\      
E &   No &    0&  Const.     &     1.2\\
\hline
\end{tabular}
\caption{
Timings $\tau$ in microseconds per timestep, $N_t$, per grid point $N_g$.
\label{timings}}
\end{center}
\end{table}

We study both laminar and turbulent regimes.
In the laminar regime we check that 
the obtained flame speed is the same as that in the one-dimensional problem. 
In the turbulent case we set the turbulent inlet  flux 
as follows. First, we consider an isothermal box 
with periodic boundary conditions. 
Initially the  density and velocity fields in the box 
are taken to be  constant.
We use a forcing function in Eq.~(\ref{eq:UU}) similar to that used in
\citet{AB01},
\EQ
f({\bm x},t)={\rm Re} \{N f_{{\bm k}(t)}
\exp [i {\bm k}(t) \cdot {\bm x}] + {\rm i} \varphi(t) \},
\EN
where ${\bm k}(t)$ is a time-dependent wavevector with
$k_{\rm f}=\langle|{\bm k}|\rangle$ being its average value that is the
chosen to be 1.5 times the minimal wavenumber that fits into the domain,
and $\varphi(t)$ is a random phase.
The prefactor $N=f_0 c_s (k_{\rm f} c_{\rm s0}/\delta t)^{1/2}$ is
chosen on dimensional grounds, $c_{\rm s0}$ is a reference sound speed,
and $f_0$ is a nondimensional factor that it chosen to regulate the
strength of the turbulence.

The simulation is run until the turbulence is statistically stationary.
This box of statistically stationary isotropic turbulence is then used as
the inlet condition for the simulation of the turbulent flame front.
The values of a two-dimensional slice from the box 
(perpendicular to the main stream) are 
used as the instantaneous inlet velocity, and
the slice is changed as a function of time to represent a real inlet.

\begin{figure*}
\centerline{\epsfig{file=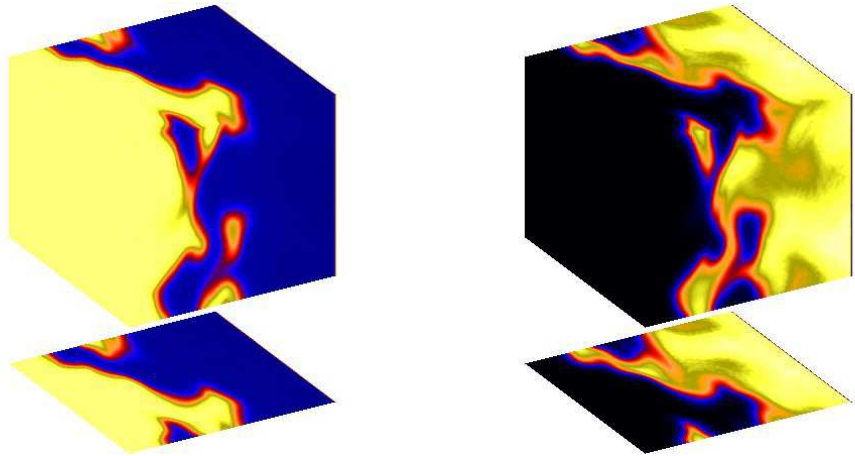,width=\textwidth}} 
\caption{From left to right instantaneous mass fractions of H$_2$ and OH  
are shown. Unburnt turbulent gas in injected on the left.}
\label{fig:llturb}
\end{figure*}

For the test case shown here the turbulent intensity is 7 times larger
than the laminar flame velocity $S_{\rm L}=10.2$ m s$^{-1}$. 
We find that for the mean inlet velocity $3S_{\rm L}$
the flame is nearly stationary inside the domain.
This indicates that the turbulent
flame velocity in this case is around $3S_{\rm L}$.
However, it is hard to determine the turbulent flame speed precisely, because
it is difficult to make the flame perfectly stationary inside the domain. 
This is partly because of the fact that between inlet and outlet
the turbulence is decaying.
Far from the inlet the turbulence is weaker than close to the inlet,
whereas the turbulent flame speed increases with the turbulent intensity.
As a result, the flame which is already far from the inlet  tends 
to move even further downstream and  the flame brush becomes broader.

\begin{figure*}
\centerline{\epsfig{file=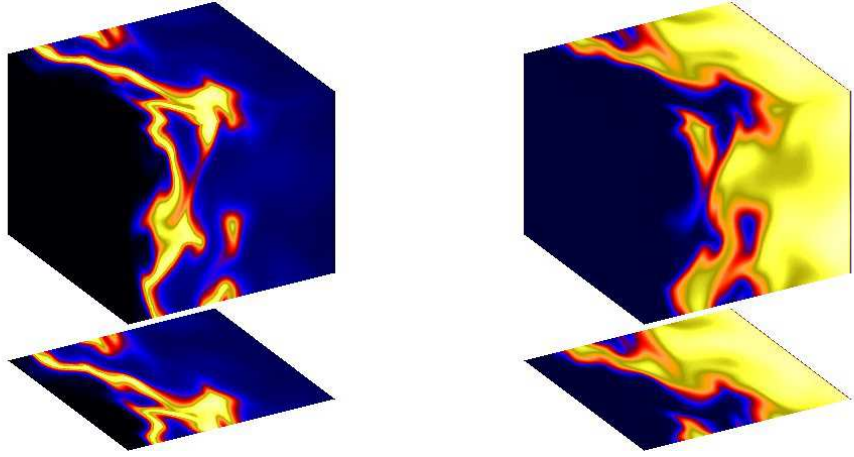,width=\textwidth}} 
\caption{As Fig.~\ref{fig:llturb}, but for the instantaneous
mass fraction of HO$_2$ on the left and temperature on the right.
}
\label{fig:llturb2}
\end{figure*}

In Fig.~\ref{fig:llturb} one can see that the H$_2$ fuel
(on the left hand side of the domain) is all consumed over the flame brush.
The thickness of the flame brush is of the order of half the box length
($2.5$~mm) and is slightly smaller than the integral scale of the turbulence. 
The mass fraction of OH is shown in the
right hand figure. It is clearly seen that OH does not
burn out after the flame, but due to the very high temperatures downstream of
the flame front the mass fraction of OH stays rather constant. For HO$_2$
the situation is however rather different and it exists only in the
neighborhood of the reaction zone of the flame (see the left
hand figure of Fig.~\ref{fig:llturb2}). This indicates that HO$_2$
might be used as an indicator of the reaction zone. In the right-hand figure
the temperature is shown to increase from 750~K to 1984~K, but the
maximum value will increase even more downstream of the box due to radical
reconnection. 

In addition, we find that the turbulence is damped behind the flame front,
and the burnt gas stream looks much more laminar there (not shown here).  
This happens because the values of temperature and hence also viscosity
of the burnt gas are much larger than those of the unburned mixture.

\subsubsection{Timings}
As DNS is very CPU intensive, it is crucial that the timings are as good
as possible. The current setup has been tested
on a single processor with different chemistry and transport data, and 
the results are presented in Table~\ref{timings}.
Run A, with the full Li mechanism and mixture averaged transport coefficients,
use the most resources, as expected. By simplifying the transport data
\citep{PO10} (Run C) \Eq{eq:Dk} is substituted by
\EQ
D_k=D_0\frac{T^n}{\rho}
\EN
and \Eq{eq:lambda} is substituted by
\EQ
\lambda=\rho c_p \kappa_0 T^n
\EN
where $n=0.7$ and $D_0=\kappa_0=2.9\times 10^{-5}$~g/(s~cm~K$^n$) leading to
a 28\% reduction in CPU consumption. Lets now turn off reactions, but still
keeping all the 13 species (Run D), and an additional 53\% reduction is achieved.

For comparison, Run~E is shown in order to see how much is gained by solving
only the Navier-Stokes equation together with the continuity equation, assuming
an isothermal medium with transport coefficients and thermodynamics such that 
all species can be neglected. It is seen that this is 20 times faster than
Run~D. This large difference is due to the fact that for Run~E only 4 equations
are solved, in contrast to the 18 equations for run D. Furthermore, and even 
more importantly, the time consuming process of determining the 
thermodynamics, such as enthalpy and heat capacity, together with the 
calculation of the viscosity, is omitted.

\subsection{Spherical flame front}
\label{sfm}
\begin{figure*}
\centerline{\epsfig{file=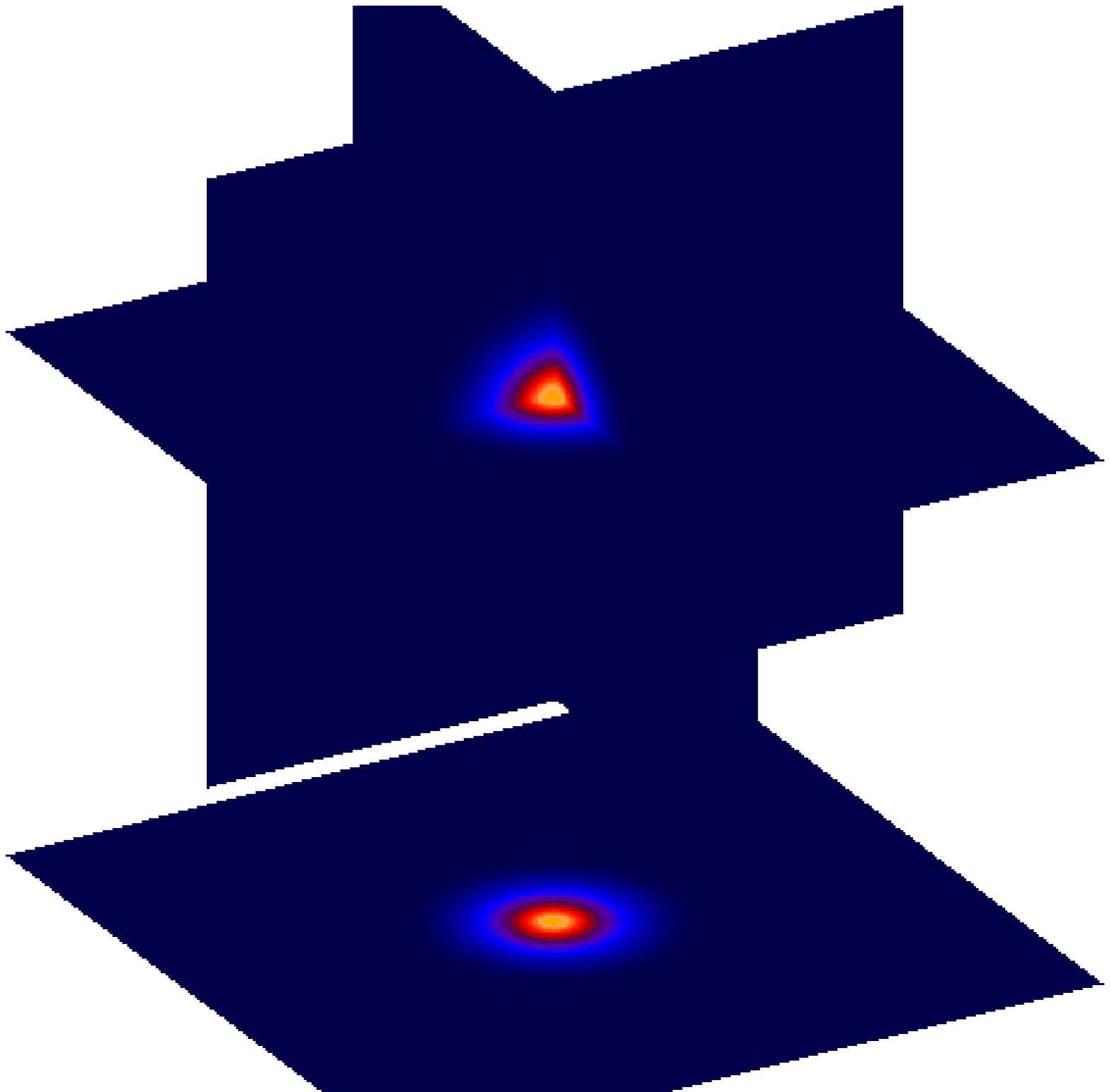,width=6cm} \epsfig{file=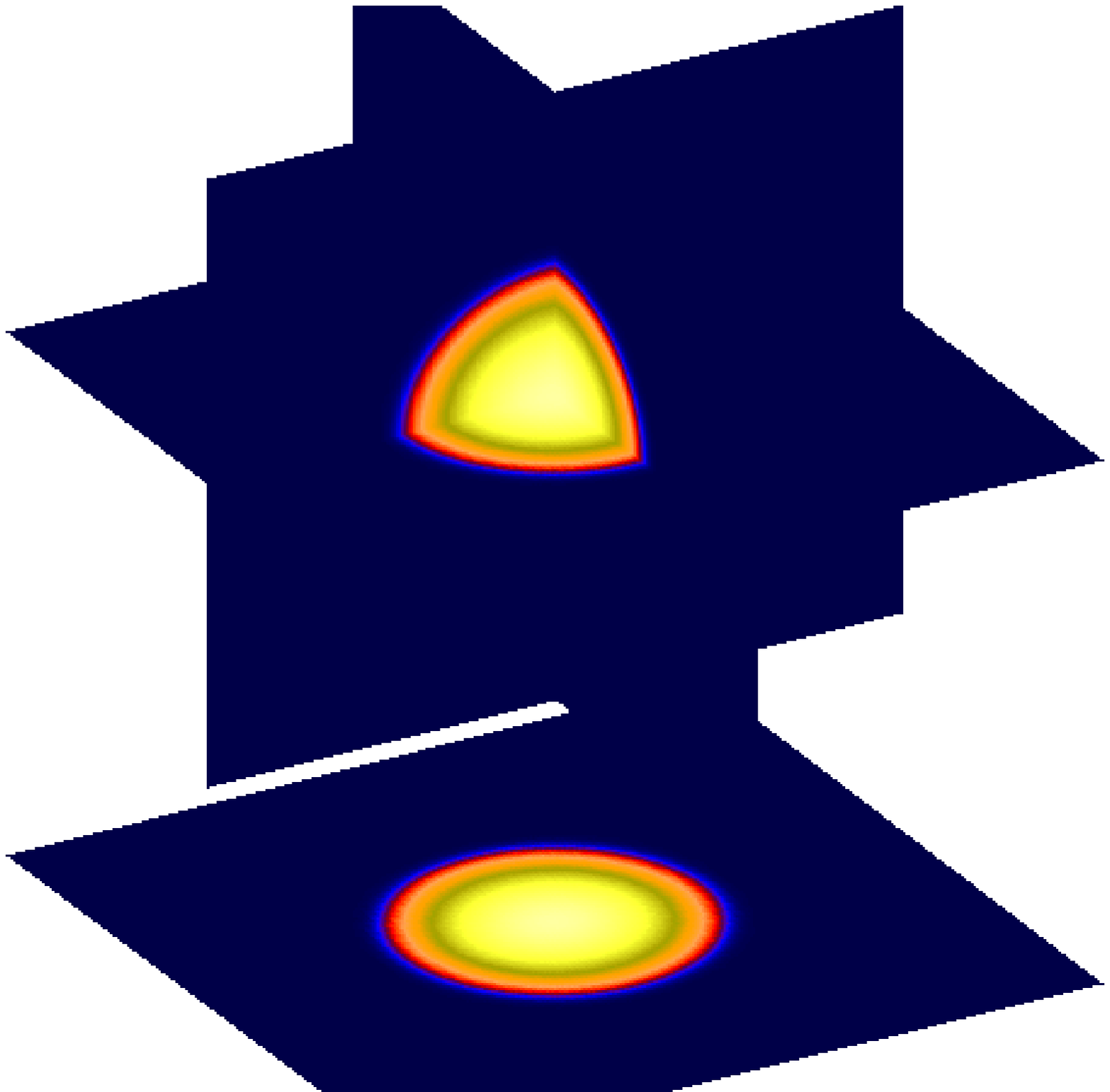,width=6cm}} 
\caption{3D images of the hot spot at moment $t=0$ s ({\it left panel}) and $t=1.2 \times 10^{-4}$ s 
({\it right panel}).}
\label{fig:img}
\end{figure*}

A study of the spherical and cylindrical flames is  important 
because these cases are useful for determining important parameters
in premixed combustion such as burning velocity, flame stretch rate,
and flame curvature.
There is a lot of numerical and experimental research in this area 
\citep{GG02}.
The most important difficulty in the numerical approach is the large
computational demand.
The typical  mesh size has to be $\delta x= 40 - 60$ $\mu$m \citep{Gr03},
i.e.\ for a cube of 3 cm$^3$ one needs about $500^3$ grid points and ideally
512 processors.

\begin{figure}
\centerline{\epsfig{file=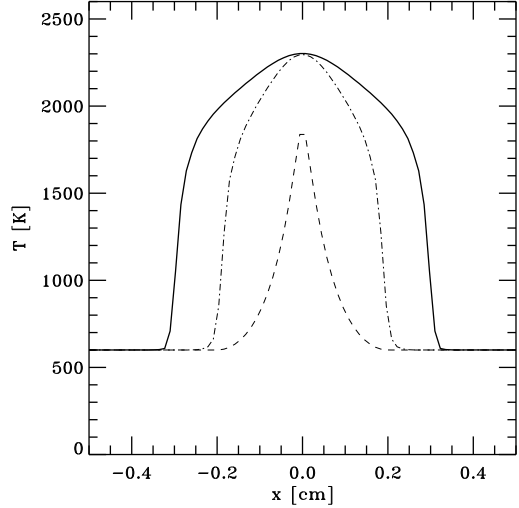,width=7 cm}} 
\caption{Temperature as a function of $x$ coordinate in the mid-plane of the box at $t=0$ s ({\it dashed curve}), $t=10^{-4}$ s ({\it dotted-dashed curve}) and $t=1.9 \times 10^{-4}$ s ({\it solid curve}).}
\label{fig:TT_x}
\end{figure}

For illustration purposes we consider a smaller cube ($1\,{\rm cm}^3$),
centered at the reference point with the hot spherical spot in its center
(see Fig.~\ref{fig:TT_x}).  The initial  hydrogen-air mixture 
with $Y_{\rm H_2}=2.4$ $\%$, $Y_{\rm O_2}=23$ $\%$
and $Y_{\rm N_2}=74.6$ $\%$ is under a pressure of $p=1$ bar.
 We use NSCBC boundary conditions, take
a grid size of (80 $\times$ 80 $\times$ 80), and 25 processors on the
Cray XT4/XT5.
The results are presented in Figs.~$\ref{fig:img}$ and $\ref{fig:TT_x}$.
The 3D images of the hot spot at the different moments (at $t=0$ s and
$t=1.2 \times 10^{-4}$ s) are presented in Fig.~$\ref{fig:img}$.  In
Fig.~$\ref{fig:TT_x}$ one sees that the gas is burned in the center
and then the flame front is expanding symmetrically in all three
directions.

This problem is also used as a good test for the fully three dimensional
NSCBC boundary conditions.  We tested the implemented NSCBC boundary
condition both for laminar and turbulent regimes.  In the laminar case
we find that due to the full NSCBC boundary conditions \citep{LD08}
the code runs well up to the moment when the flame front comes to the
domain boundaries.  In the turbulent regime the problems appear near
the corners and edges of the domain because of the eddies at the
boundaries.  We avoid such a problem by using buffer (or sponge) zones (for
details see \citep{BB08}).  We add the term to the right-hand side of
the momentum equation
\begin{eqnarray}
\frac{{\rm D} V_{i}^{j}}{{\rm D}
t}=...-\frac{V_{i}^j-V_{{\rm ref},i}}{\tau} \zeta(x_i),\quad j=1, ...,
N_i, \label{eq:bz}
\end{eqnarray}
where $j$ denotes the meshpoint and
$N_i$ is total number of grid points in the $i$ direction,
and dots indicate the presence of terms that where already specified
in equation (\ref{eq:UU}), $\zeta(x_i)$ is equal to zero everywhere except
in the buffer zones where it is equal to unity. 
The length of the buffer zone is 10 $\%$ of the domain,
and we choose $\tau=5 \delta t$ and $V_{{\rm ref},i}=0$.

\section{Conclusions}

In this paper we have presented a high-order public domain code for direct
numerical simulation of compressible flows with detailed chemical reactions.
The {\sc Pencil Code} provides sixth-order spatial
accuracy in the simple one-step reaction case, and fifth order accuracy in
the case where upwinding for density advection is necessary.
For validation purposes we compare
our results with the Chemkin tool for 0D and 1D test problems, and show
that they are in good agreement. Finally, we calculate the flame
speed in 3D both in laminar and turbulent cases.

The code is well suited for considering also more complicated
reaction schemes such as methane combustion.
Furthermore, it is straightforward to consider the interaction with
additional chemicals such as nitrogen and to follow the production of
NOx gases.
In particular, it is important to consider combustion in the presence
of steam.
This is well known to lead to a reduction of NOx gases.
Combustion in the presence of more complicated boundary conditions
involving, for example, smaller inlet geometries has also been considered.
Some of these cases, including those with a turbulent inlet,
are available among the many sample cases that come with the code.
For the benefit of the community, it is advantageous if prospective
contributers to the code ask one of the code owners listed on
\url{http://pencil-code.googlecode.com/} to obtain permission
as a committer.

\section*{Acknowledgments}

The authors thank Professor Chung K.\ Law and Mr.\ Fan Yang for
providing their data on the flame speed velocity.
We have benefitted from discussions with X.-S. Bai, V.\ Sabelnikov,
N.\ Swaminathan, U.\ Paschereit and other participants of the Nordita
Programme on Turbulent Combustion in 2010.
We acknowledge the allocation of computing resources provided by the
Center for Scientific Computing in Finland.
This work was supported by the Academy of Finland and
the Magnus Ehrnrooth Foundation (NB).
The research leading to these results has received funding from the European 
Community's Seventh Framework Programme (FP7/2007-2013) under grant 
agreement nr 211971 (The DECARBit project) (NELH).
This work was also supported in part by
the European Research Council under the AstroDyn Research Project No.\ 227952
and the Swedish Research Council Grant No.\ 621-2007-4064 (AB).


\newcommand{\jcp}[3]{ #1, {J.\ Comput.\ Phys.,} {#2}, #3}
\newcommand{\ymn}[3]{ #1, {Monthly Notices Roy.\ Astron.\ Soc.,} {#2}, #3}
\newcommand{\yapj}[3]{ #1, {Astrophys.\ J.,} {#2}, #3}
\newcommand{\yjour}[4]{ #1, {#2}, {#3}, #4}
\newcommand{\yproc}[5]{ #1, in {#3}, ed.\ #4 (#5), #2}

\vfill\bigskip\noindent\tiny\begin{verbatim}
$Header: /var/cvs/brandenb/f90/pencil-natalia/chemistry/tex/method/paper.tex,v 1.104 2010-05-28 11:53:45 nbabkovs Exp $
\end{verbatim}

\end{document}